\documentclass[12pt] {article}		
\usepackage{makeidx}
\usepackage{epsfig}
\usepackage[dvips]{color}
\usepackage{latexsym}

\normalsize
\newcommand{\cleqn}{\setcounter{equation}{0}}

\newcommand{\be}{\begin{equation}}
\newcommand{\ee}{\end{equation}}
\newcommand{\bea}{\begin{eqnarray}}
\newcommand{\eea}{\end{eqnarray}}
\newcommand{\gsim}{\hbox{ \raise3pt\hbox to 0pt{$>$}\raise-3pt\hbox{$\sim$} }}
\newcommand{\lsim}{\hbox{ \raise3pt\hbox to 0pt{$<$}\raise-3pt\hbox{$\sim$} }}
\newcommand{\mathbold}[1]{\mbox{\boldmath $#1$}}

\newcommand{\figdir}{figs}
\begin{document}
\begin{titlepage}

\hfill
\begin{minipage}[t]{2.0in}
\begin{flushleft}
\bf
KEK Preprint 2000-113 \\
KUDP 2000-001 \\
TUAT-HEP 2000-04 \\
KINDAI-HEP 00-01 \\
November 2000 \\
H \\
\end{flushleft} 
\end{minipage}

\vspace*{1.0cm}

\begin{center}
\Large \bf
Lorentz Angle Measurement\\
for \protect\boldmath${\rm CO}_2$/Isobutane Gas Mixtures
\end{center}

\vspace*{1.0cm}

\begin{center}
\large 
	$K.Hoshina^b$, $K.Fujii^a$, $N.Khalatyan^c$,
	$O.Nitoh^b$, $H.Okuno^a$,\\ 
	$Y.Kato^e$,
	$M.Kobayashi^a$, $Y.Kurihara^a$, 
	$H.Kuroiwa^b$, $Y.Nakamura^b$,\\
	$K.Sakieda^b$, $Y.Suzuki^b$,
	$T.Watanabe^d$\\
\end{center}

\vskip 0.7cm

\begin{center}
$^a$ High Energy Accelerator Research Organization(KEK),
Tsukuba, 305-0801, Japan\\ 
$^b$Tokyo University of Agriculture and Technology,
Tokyo 184-8588, Japan\\
$^c$Institute of applied physics, University of Tsukuba, 305-8573, Japan\\
$^d$Kogakuin University, Tokyo 163-8677, Japan \\
$^e$Kinki University, Osaka, 577-0085, Japan
\end{center}


\vspace*{3cm}

\begin{abstract}

We have developed a Lorentz angle measurement system for
cool gas mixtures in the course of our R\&D for a
proposed JLC central drift chamber (JLC-CDC).
The measurement system is characterized by the use of two
laser beams to produce primary electrons
and flash ADCs to read their signals simultaneously.
With this new system, we have measured Lorentz angles
for ${\rm CO}_2$/isobutane gas mixtures with different 
proportions (95:5, 90:10, and 85:15), 
varying drift field from $0.6$ to $2.0~{\rm kV/cm}$
and magnetic field up to $1.5~{\rm T}$.
The results of the measurement are
in good agreement with GARFIELD/MAGBOLTZ simulations.
\vspace*{5mm}
\begin{flushleft}
{\it Keywords:} Lorentz angle; ${\rm CO}_2$/isobutane; Drift chamber; JLC
\end{flushleft}

\end{abstract}

\end{titlepage}



\section{Introduction}
\cleqn

In order to make maximum use of the full physics potential
of a future linear collider such as JLC~\cite{Ref:jlc},
it is highly desirable that
its detector system allows reconstruction of
final states in terms of known standard-model partons:
charged leptons, quarks, gauge bosons, and
neutrinos as missing momenta.
Among these partons,
lighter quarks ($u, d, s, c$, and $b$) and gluons
can be detected as jets with or without secondary or
tertiary vertices, 
while heavier partons ($t$, $W^\pm$, and $Z$) 
can be recognized through calculation of
jet invariant masses.
The parton reconstruction based on the jet
invariant-mass method requires not only efficient
and high-resolution tracking of charged particles
in a jetty environment, but also good track-cluster
matching to achieve the best energy flow measurements.
The central tracker should thus be capable of
measuring individual charged particles in jets
with high efficiency and high momentum resolution,
as well as good angular resolution.
In order to satisfy these demands, we have proposed,
as a candidate JLC central tracker~\footnote{
Detailed design parameters of the JLC-CDC can be found
in our previous papers \cite{Ref:cosmic},\cite{Ref:stereo}.
},
a large cylindrical drift chamber
with mini-jet cells 
filled with a slow gas mixture,
${\rm CO}_2$/isobutane(90:10).

The Lorentz angle, which is the angle between 
the drift direction of electrons under the influence
of magnetic field and the direction of electric field,
is one of the key parameters to determine
the jet cell design.
The existence of the magnetic field tilts 
drift lines by the Lorentz angle with respect to the
electric field direction.
If this angle is too big, drift lines for the edge
wires hit top or bottom walls of the drift cell,
thereby leading us to loss of detection efficiency.

\begin{figure}[htb]
\centerline{
\begin{minipage}[htb]{12cm}
\centerline{
\epsfxsize=10cm  
\epsfbox{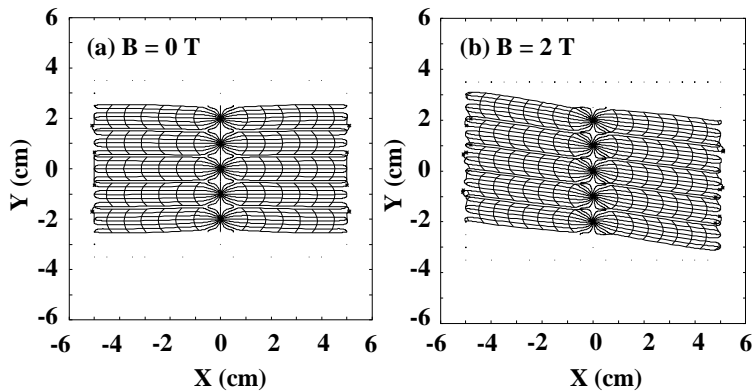}
}
\caption[Fig:drift-magn]
{\small \label{Fig:drift-magn}
GARFIELD results of electron drift lines and iso-chrones 
in the JLC-CDC for the ${\rm CO}_2$/isobutane(90:10) gas mixture.
}
\end{minipage}
}
\end{figure}

Fig.~\ref{Fig:drift-magn} shows electron drift lines in a jet cell
calculated by the chamber simulation program GARFIELD\footnote{
GARFIELD provides interfaces to different program
packages for calculations of gas properties such as
drift velocity, diffusion coefficient, Lorentz angle, etc.
Among such interfaces, we used one for MAGBOLTZ, 
which numerically solves Boltzmann's transport equation for
drifting electrons.
Unless otherwise specified,
the latest version of GARFIELD as of Sep., 2000 (version 7)
uses MAGBOLTZ-2\cite{Ref:MAGBOLTZ2}, which exploits a Monte Carlo integration
technique for the solution, whereas the older
versions use MAGBOLTZ-1\cite{Ref:MAGBOLTZ} based on an analytic formulation.
The two versions yielded consistent results for our gas mixtures at
some representative electric and magnetic field values.
We thus decided to use the analytic version (version 1.16),
since it was impracticable, from computing time point of view,
to achieve required accuracy for comparison with all of our data points
using the Monte Carlo version (version 2.2)
(see Subsection~\ref{Sub:Results}).
}~\cite{Ref:GARFIELD}
(a) without and (b) with the magnetic field. 
The results indicates that the drift lines are
completely contained in the cell
at least up to a magnetic field of $2~{\rm T}$.
There is, however, no systematic experimental data published
for the Lorentz angles of ${\rm CO}_2$/isobutane gas mixtures,
which makes difficult for us
to test the reliability of our calculations 
using the GARFIELD program.
In order to confirm the results of the GARFIELD calculations
and to verify the validity of our cell design, 
we have developed a Lorentz angle measurement system 
for cool gas mixtures, which have small Lorentz angles,
and carried out systematic measurement for 
${\rm CO}_2$/isobutane gas mixtures.
The measurement system is characterized by the use of two
laser beams to produce primary electrons
and flash ADCs to read their signals simultaneously.

In this paper we describe the measurement system in detail,
present the obtained Lorentz angle data, 
and compare them with GARFIELD/MAGBOLTZ simulations.
The paper is organized as follows:
Section 2 introduces some basic formulae relevant to
the Lorentz angle measurement, and
outlines the design philosophy of our measurement system. 
Section 3 then describes its hardware aspects,
including optical setup, the test chamber geometry,
and data acquisition.
Section 4 sketches the analysis procedure and
presents the Lorentz angle results
together with corresponding GARFIELD/MAGBOLTZ predictions.
In Section 5,
these results are compared with the basic formulae
given in Section 2,
and their implications are discussed from the view point of
application to the JLC-CDC.
Finally Section 6 summarizes the results and concludes
the paper.

%
%

\section{Principle of Lorentz Angle Measurement}
\cleqn
\subsection{A Quick Theoretical Review of the Lorentz Angle}
\label{Sub:TheoreticalReview}

In order to clarify the principal parameters
that control the Lorentz angle of drifting electrons,
let us start our discussion with deriving  
a simple expression for the drift velocity vector
in the presence of electric and magnetic fields, $\mathbold{E}$ and 
$\mathbold{B}$~\cite{KL87}.
The equation of motion for a drifting electron 
under the influence of the $\mathbold{E}$ and $\mathbold{B}$ fields is given by
\begin{equation}
  m \frac{{\rm d} \mathbold{v}}{{\rm d} t} 
  = -e \left( \mathbold{E} + \mathbold{v} \times \mathbold{B} \right) + m \mathbold{A}(t),
\label{Eq:motion}
\end{equation}
where $m$ and $-e$ are the electron mass and charge, respectively, 
while $\mathbold{A}(t)$ is the time-dependent acceleration, 
or more appropriately deceleration,
due to stochastic force
(friction) from the surrounding gas molecules.
Since we are only interested in a steady state where the electron
drifts at a constant drift velocity:
$\mathbold{v}_D \equiv \left< \mathbold{v} \right>$,
the time-average of the left-hand side of Eq.(\ref{Eq:motion})
must vanish.
The crucial step now is to assume that
the time average of $\mathbold{A}(t)$ is
anti-parallel with $\left< \mathbold{v} \right>$:
\begin{equation}
  \left< \mathbold{A}(t) \right> = - \frac{\left< \mathbold{v} \right>}{\tau},
\label{Eq:avat}
\end{equation}
where the constant $\tau$ has the dimension of time
and is called characteristic time 
which can be related to the average time interval
between collisions.
The time average of Eq.(\ref{Eq:motion}) then becomes
\begin{equation}
\label{Eq:taeqm} 
   0 = -e \left( \mathbold{E} + \mathbold{v}_D \times \mathbold{B} \right) 
       - \frac{m}{\tau} \mathbold{v}_D,
\end{equation}
which can be cast into the form:
\begin{equation}
  \left( \frac{m}{\tau} - e \mathbold{B} \times \right) \mathbold{v}_D = -e \mathbold{E}.
\end{equation}
Rewriting this as a matrix equation and inverting
the matrix, we can solve this for $\mathbold{v}_D$ and obtain
an expression valid for general field configurations:
\begin{equation}
   \mathbold{v}_D = 
   \left(
     \frac{ - \mu E }
          { 1 + (\omega\tau)^2 }
   \right)
   \left(
      \mathbold{\hat{E}} 
      - (\omega\tau) \left[ \mathbold{\hat{E}} \times \mathbold{\hat{B}} \right]
      + (\omega\tau)^2 \left( \mathbold{\hat{E}} \cdot \mathbold{\hat{B}} \right) \mathbold{\hat{B}}
   \right),
\label{Eq:vdsol}
\end{equation}
where $\mathbold{\hat{E}}$ and $\mathbold{\hat{B}}$
denote unit vectors along $\mathbold{E}$ and $\mathbold{B}$, respectively,
while $\mu$ and $\omega$ are given by
\begin{equation}
   \mu \equiv \left( \frac{e}{m} \right) \tau, ~~~
   \omega \equiv \left( \frac{e}{m} \right) B
\label{Eq:defmuomega}
\end{equation}
and called mobility and cyclotron frequency, respectively.
This solution consists of three components:
one parallel
with $\mathbold{E}$, another parallel with $\mathbold{B}$,
and the rest orthogonal to both $\mathbold{E}$ and $\mathbold{B}$.
When $\omega \tau \simeq 0$, which means 
low magnetic field or short time
interval between collisions,
$\mathbold{v}_D$ tends to the $\mathbold{E}$ field direction.
On the other hand, when $\omega \tau \gg 1$,
the electron curls up into the $\mathbold{B}$ field direction,
as long as $\mathbold{E} \cdot \mathbold{B} \ne 0$.

It is easy now to derive from Eq.(\ref{Eq:vdsol})
general expressions for the magnitude
of $\mathbold{v}_D$,
$v_D = \left| \mathbold{v}_D \right|$:
\begin{eqnarray}
   v_D
   = \mu E ~
   \sqrt{
     \frac
     {1 + \left( \mathbold{\hat{E}} \cdot \mathbold{\hat{B}} \right)^2 (\omega\tau)^2}
     {1 + (\omega\tau)^2},
   }
\label{Eq:vdmag}
\end{eqnarray}                                                                  
and the magnitudes of parallel and perpendicular components of $\mathbold{v}_D$,
$v_{D\parallel} = \left| \mathbold{v}_{D\parallel} \right|$ and 
$v_{D\perp} = \left| \mathbold{v}_{D\perp} \right|$, to the $\mathbold{E}$ field:
\begin{eqnarray}
   v_{D \parallel} 
   & = & \mu E ~
     \frac
     {1 + \left( \mathbold{\hat{E}} \cdot \mathbold{\hat{B}} \right)^2 (\omega\tau)^2}
     {1 + (\omega\tau)^2}
\label{Eq:vdpara}
\end{eqnarray}
and
\begin{eqnarray}
   v_{D \perp} 
   & = & \mu E ~ \omega \tau ~
     \frac
     { 
       \left| \mathbold{\hat{E}} \times \mathbold{\hat{B}} \right|
       \sqrt{
       	  1 + \left( \mathbold{\hat{E}} \cdot \mathbold{\hat{B}} \right)^2 (\omega\tau)^2
       }
     }
     {1 + (\omega\tau)^2}.
\label{Eq:vdperp}
\end{eqnarray}

Dividing Eq.(\ref{Eq:vdperp}) by Eq.(\ref{Eq:vdpara}), we thus arrive
at the following equation for the Lorentz angle $\alpha$:
\begin{eqnarray}
   \tan \alpha 
   & = &
     \frac
     { \left| \mathbold{\hat{E}} \times \mathbold{\hat{B}} \right| }
     {
     	\sqrt{
     	  1 + \left( \mathbold{\hat{E}} \cdot \mathbold{\hat{B}} \right)^2(\omega\tau)^2
     	}
     }
     ~ \omega \tau.
\label{Eq:tana}
\end{eqnarray}
For completeness, we show below the formula for angle $\beta$
that is the angle between $\mathbold{v}_{D\perp}$ and 
$\mathbold{\hat{E}} \times \mathbold{\hat{B}}$:
\begin{eqnarray}
   \tan\beta
   & = &
        \left( \mathbold{\hat{E}} \cdot \mathbold{\hat{B}} \right)
     ~ \omega \tau.
\label{Eq:tanb}
\end{eqnarray}

If $\mathbold{E}$ and $\mathbold{B}$ are mutually orthogonal, 
$\mathbold{\hat{E}} \cdot \mathbold{\hat{B}} = 0$,
as in our cell design,
we then have
$\left| \mathbold{\hat{E}} \times \mathbold{\hat{B}} \right| = 1$,
which leads us to a simple set of equations:
\begin{equation}
v_D = \frac{\mu E}{\sqrt{1+(\omega\tau)^2}}
\label{Eq:vdsimple}
\end{equation}
and
\begin{equation}
   \tan \alpha = \omega \tau,
\label{Eq:tanasimple}
\end{equation}
while $\tan\beta = 0$ meaning that $\mathbold{v}_{D\perp}$
is perpendicular to both $\mathbold{E}$ and $\mathbold{B}$.
It is remarkable that the Lorentz angle is
determined by just two parameters $\omega$ and $\tau$,
where only $\tau$ reflects the complexity 
involved in the electron drift process.

Using Eqs.(\ref{Eq:defmuomega}) and (\ref{Eq:vdsimple})
and assuming that $\tau$ is independent of $B$,
we can rewrite Eq.(\ref{Eq:tanasimple}) in the following form:
\begin{equation}
   \tan \alpha = \left( \frac{B}{E} \right) v_D^0,
\label{Eq:tana-psi=1}
\end{equation}
where $v_D^0 = v_D(B = 0)$.
On the other hand,
the Lorentz angle is often parameterized~\cite{Ref:Huxley}\cite{Ref:kunst} empirically as
\begin{equation}
   \tan \alpha = \psi \left( \frac{B}{E} \right) v_D^0,
\label{Eq:tanA}
\end{equation}
where the dimensionless factor $\psi$ is called a
magnetic deflection coefficient.
The above formulation thus predicts $\psi = 1$.
We shall return to this point later in Section~\ref{Sec:discussion}.

\subsection{Theoretical Expectation for \protect\boldmath${\rm CO_2}$/isobutane Gas Mixtures}
\label{Sub:TheoreticalExpectation}

The formulae presented above allow us to estimate Lorentz angles
for ${\rm CO}_2$/isobutane gas mixtures for a given set of
$\mathbold{E}$ and $\mathbold{B}$ fields.
For instance, Eq.(\ref{Eq:tana-psi=1}) predicts that
the Lorentz angle is proportional to the drift velocity.
Because of low drift velocity, any cool gas mixture such as
those having ${\rm CO}_2$ as their main component must have
a small Lorentz angle.
For example, our candidate gas mixture for the JLC-CDC,
${\rm CO}_2$/isobutane (90:10), 
has a drift velocity of 
$v_D^0 = 0.78~{\rm cm}/\mu{\rm s} = 7.8~{\rm \mu m/}{\rm ns}$
at $E = 1.0~{\rm kV/cm} = 100~{\rm kV/m}$.
At $B = 1~{\rm T}$ this drift velocity can be
translated into a Lorentz angle of
\begin{equation}
   \tan\alpha = \left( \frac{B}{E} \right) v_D^0
   = \frac{1~[{\rm T}]}{100~[{\rm kV/m}]} \times 7.8~[{\rm \mu m/}{\rm ns}]
   = 0.078.
\end{equation}
This angle corresponds to a transverse displacement of
about $1.5~{\rm mm}$ for a drift distance of $2~{\rm cm}$.
Since Lorentz angle measurements are mostly based on 
the determination of this transverse displacement,
high resolution and low systematic error are key
requirements for its measurement.
For instance, an error of $50~\mu{\rm m}$ on the displacement,
already amounts to a relative error of about $3~\%$
on $\tan\alpha$ for the aforementioned sample case.

\subsection{Basic Principle of the Measurement}
\label{Sub:BasicPrincipleofMes.}
 
Basic principle of the Lorentz angle determination
is as follows.
Consider a cluster of electrons
drifting along a uniform 
electric field ($\mathbold{E}$)
applied in the direction of the $z$-axis.
The presence of a uniform 
magnetic field ($\mathbold{B}$) 
in the direction of the $y$-axis
deflects the drift direction
towards the $x$-axis.
The transverse displacement ($\Delta x$)
after drifting over a given distance ($\Delta z$)
is a direct measure of the Lorentz angle:
\begin{equation}
  \tan\alpha = \frac{\Delta x}{\Delta z}.
\end{equation}
In principle,
we thus need to measure just $\Delta x$ and $\Delta z$,
provided that $\mathbold{E}$ and $\mathbold{B}$ are
uniform, orthogonal, stable, and known.

In practice, the uniform electric field is
provided by field-shaping plates in a drift chamber.
The measurement of the transverse
displacement, $\Delta x$, can not be made without interfering with 
the drifting electron cluster:
we need to lead the cluster to a gas amplification region,
where the $\mathbold{E}$ field varies quite drastically.
The drift region where the $\mathbold{E}$ field 
is required to be uniform thus has to be separated
from the amplification region with a separating slit.
The complexity involved near the slit and in the
amplification region has to be somehow canceled.
On the other hand, the measurement of $\Delta z$
seems rather straightforward,
since we can control the starting point
of the drifting electron cluster, for instance,
by controlling the position of the injection point
of a laser beam that creates the cluster.
It is, however, non-trivial to measure the laser beam
position at the ionization point.
In the following subsection, we will explain
how to avoid these potential problems.

\subsection{Our Measurement System}

In order to cancel the effects of the complicated processes
due to the non-uniform
field near the slit and in the amplification region,
it is a common practice\cite{Ref:becker,Ref:vavra}
to create primary electron clusters
at different $z$ points ($z_1$ and $z_2$),
by moving a laser beam in the $z$ direction,
and measure their arrival points ($x_1$ and $x_2$) at 
detecting electrodes.
In our case, the electron clusters pass through the slit
and then arrive at a sense anode wire strung perpendicular to
both $\mathbold{E}$ and $\mathbold{B}$.
The avalanche locations along the anode wire,
corresponding to $x_1$ and $x_2$, 
are then determined by the charge centroid method
with a set of cathode pads behind it.
We can then take the differences
${\Delta x = x_2 - x_1}$ and
${\Delta z = z_2 - z_1}$
to calculate the Lorentz angle.
The method assumes that the two starting positions
have the same $x$ coordinate, or if they are different,
they can be calibrated by measuring ${\Delta x}$ 
for no magnetic field.

As shown in Subsection \ref{Sub:TheoreticalExpectation},
however,
we expect only small transverse displacements for 
${\rm CO}_2$-based gas mixtures.
Any small systematic change or instability of
the laser beam position is a potential source of 
significant error.
Instead of moving the laser beam, we thus decided to split it
into two parallel beams and inject them simultaneously
at different $z$ positions.
Since the relative distance of the two laser beams is
fixed by the splitter and since the relative distance
is the only quantity that actually matters,
we can cancel most systematics due to systematic shift
or instability of the laser beam positions relative to the chamber.


\begin{figure}[htb]
\centerline{
\begin{minipage}[htb]{12cm}
\centerline{
\epsfxsize=6cm
\epsfbox{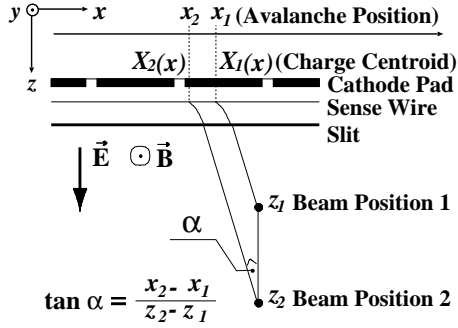}
}
\caption[Fig:principle]
{\small \label{Fig:principle}
Principle of the Lorentz angle measurement.
}
\end{minipage}
}
\end{figure}

\begin{figure}[htb]
\centerline{
\begin{minipage}[htb]{12cm}
\centerline{
\epsfxsize=10cm
\epsfbox{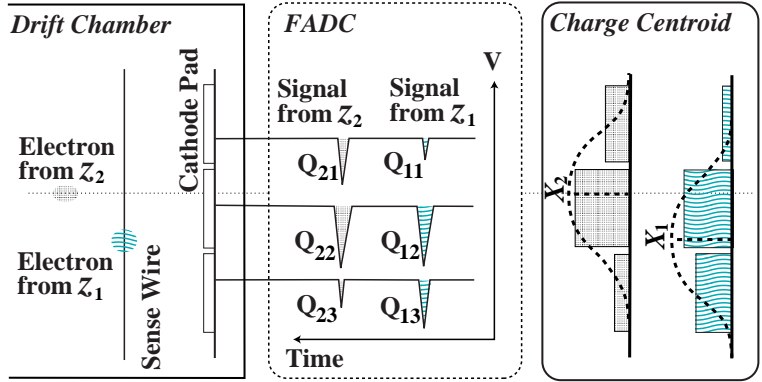}
}
\caption[Fig:DAQ]
{\small \label{Fig:DAQ}
Data acquisition scheme.
}
\end{minipage}
}
\end{figure}

Now that two laser beam pulses are injected simultaneously,
and since the transverse displacement of the electron clusters
is small for our gas mixtures, their induced charge
distributions on the cathode pads overlap in space.
They are, however, separable in time, since their 
corresponding drift distances differ.
We thus use flash ADCs to record
the signal amplitudes as a function of time, and
reconstruct individual induced charge
distributions on the cathode pads.
Fig.~\ref{Fig:DAQ} illustrates how our
DAQ system enables us to
separately measure the individual charge centroids
of the two electron clusters.
%
%

The advantages of this two-beam system can be summarized as follows:
\begin{itemize}
\item
We can eliminate most of the effects of spatial jitters
of the laser beams with respect to the chamber.
Figure \ref{Fig:timedepend} demonstrates how powerful
this scheme is in the case of occasional instability
we experienced for higher magnetic fields.
The oscillations of the two arrival points are
coherent and therefore can be canceled
by taking their difference.
%
\begin{figure}[htb]
\centerline{
\begin{minipage}[htb]{12cm}
\centerline{
\epsfxsize=10cm
\epsfbox{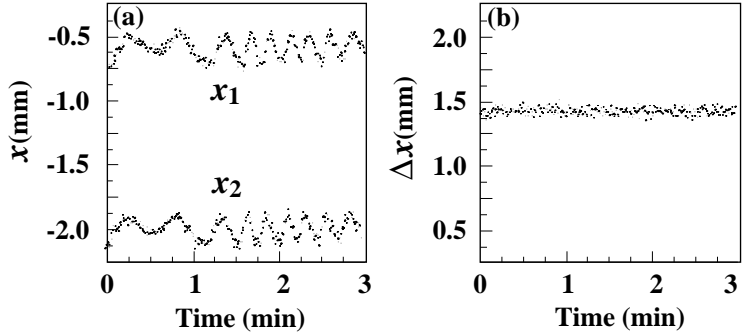}
}
\caption[Fig:timedepend]
{\small \label{Fig:timedepend}
Example of occasional instability observed at $B = 1.2~{\rm T}$
in the time-dependence of (a) the arrival points of 
two electron clusters
($x_1$ and $x_2$).
The oscillation is absent from
their difference as shown in (b).
}
\end{minipage}
}
\end{figure}
%
\item
We can calibrate the laser beam distance
($\Delta z = z_2 - z_1$) 
by using the recorded time interval of 
the two anode pulses,
provided that the drift velocity has been 
measured separately.
It should be stressed here that this measures the
laser beam distance at the very points of ionization. 
\item
Baselines or pedestals for the cathode pad signals
can be monitored on an event-by-event basis.
\item
The two arrival points are measured with the same
set of cathode pads, which is a virtue of slow gas
mixtures with small Lorentz angles, thereby
reducing systematic errors due to channel-to-channel
variation of the readout electronics.
\end{itemize}
There are, however, some drawbacks: 
among them potentially the most serious problem, which deserves
special attention, is the space-charge effect
of the avalanche formed by the first electron
cluster on the second one.
We will discuss this in Section~\ref{Sub:Analysis}.

%
%

\section{Experimental Setup}
\begin{figure}[htb]
\centerline{
\begin{minipage}[htb]{12cm}
\centerline{
\epsfxsize=12cm
\epsfbox{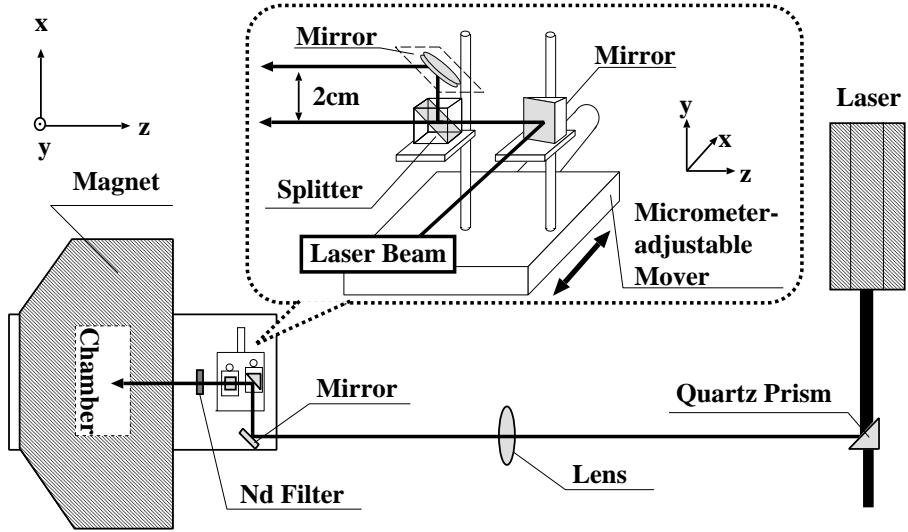}
}
\caption[Fig:setup]
{\small \label{Fig:setup}
Schematic view of the experimental setup, where
our coordinate axes are also defined.
}
\end{minipage}
}
\end{figure}

\cleqn
In this section, we describe how we realized
the idea presented above in a real hardware form.
Our measurement system can be divided into
three major parts: (1) a laser beam system including
a laser pulse generator and an optical system
to prepare two parallel laser beams,
(2) a cathode-readout drift chamber and its gas system,
together with a dipole magnet to supply
a magnetic field in the chamber's drift region, and
(3) a data acquisition system featuring flash ADCs
to record the time-profiles of signals from the cathode pads.
A schematic view of the setup 
is shown in Fig.~\ref{Fig:setup}, and in what follows,
we will describe these three parts in more detail.
%

\subsection{Laser Beam System}
\label{Sub:LaserBeamSystem}

The laser beam system can be subdivided into
a laser pulse generator and an optical system
including a beam splitter.
An Nd-YAG laser
with a wave length of $266~{\rm nm}$ 
generates beam pulses having energies up to $50~{\rm mJ}$ and
a duration of about $5~{\rm nsec}$
at a frequency of $10~{\rm Hz}$.
The intensity of the laser beam is reduced to $3~\%$ 
while reflected by a quartz prism.
It is then focused with a lens having a focal length of
$3~{\rm m}$, and reflected twice with a pair of $45^\circ$ mirrors,
the second of which is mounted on a micrometer-adjustable mover
and allows us to move the laser beam along the $x$-axis.
The reflected beam is split into two beams 
by a cubic splitter consisting of two prisms
glued together\cite{Ref:sigma-koki}:
a part of the beam is reflected upward at the boundary,
while the rest passes straight through.
The upward beam is reflected again by yet another $45^\circ$
mirror placed $2~{\rm cm}$ apart from the splitter, and
goes in parallel with the straight-through beam.
In order to control the relative intensities of
the split beams, we put an ND filter in the way
of the pass-through beam.
The two laser beams are now ready for disposal
at the chamber.

\subsection{Drift Chamber}

\begin{figure}[htb]
\begin{minipage}[htb]{6cm}
\centerline{
\epsfxsize=5cm 
\epsfbox{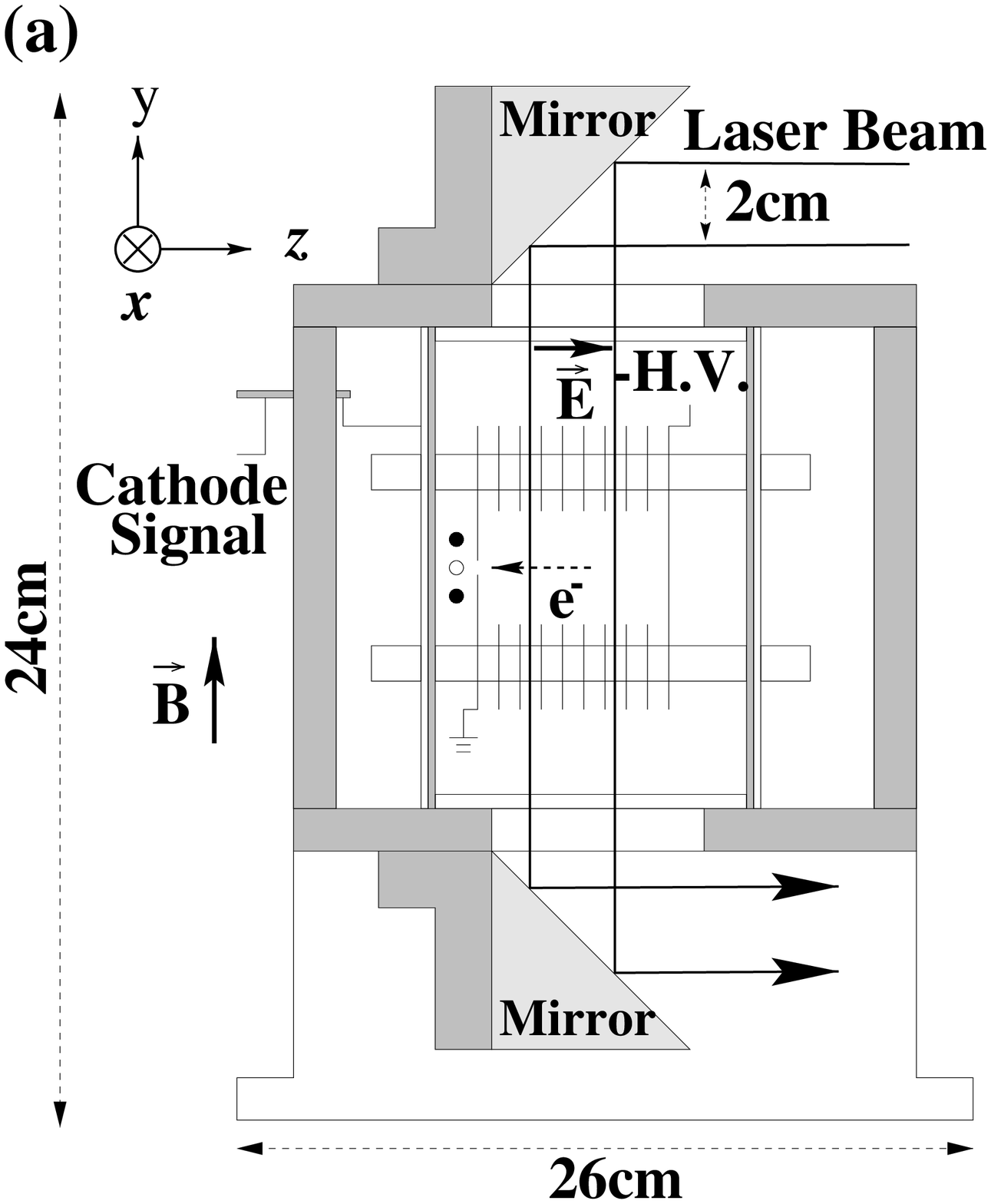}
}
\end{minipage}
\hfill
\begin{minipage}[htb]{9cm}
\centerline{
\epsfxsize=9cm 
\epsfbox{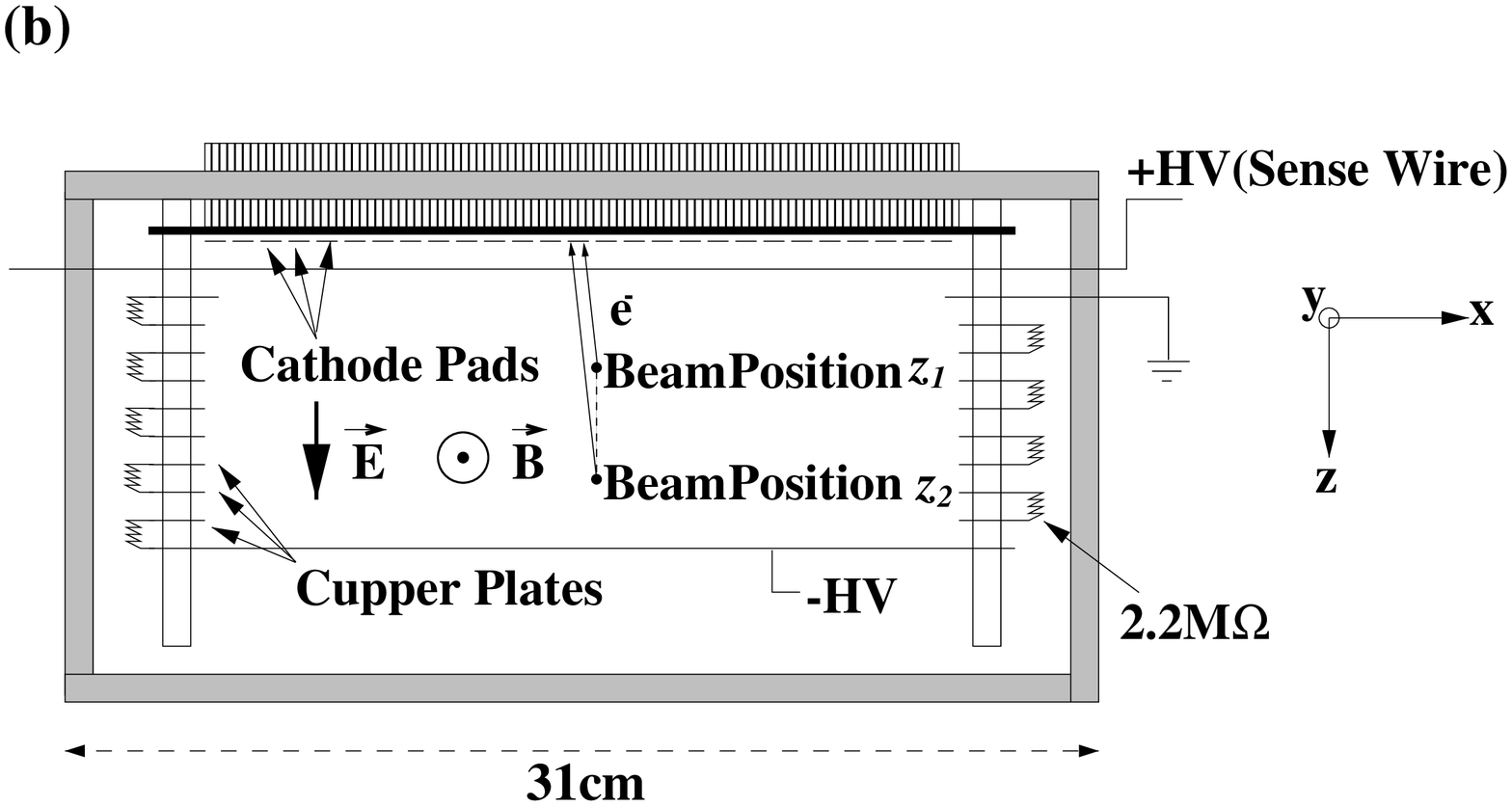}
}
\end{minipage}
\caption[Fig:chamber]
{\small \label{Fig:chamber}
(a) side- and (b) top-views of 
the drift chamber.
}
\end{figure}

Our Lorentz angle measurement chamber\cite{OK98} has a structure
depicted in Figs.~\ref{Fig:chamber}-(a) and (b)
and is placed in a uniform magnetic field up to
$1.5~{\rm T}$ provided by a dipole magnet named
USHIWAKA in the KEK-PS $\pi 2$ beam line.
The magnet has an aperture of $54 \times 64~{\rm cm}^2$
and a pole tip length of $30~{\rm cm}$ and is
capable of providing a maximum field of $1.6~{\rm T}$
with a uniformity much better than $1\%$ over the
region occupied by the chamber. 
%
%
As seen in Fig.~\ref{Fig:chamber}-(a),
the two laser beams are reflected vertically down,
which is the direction of the magnetic field,
by a $45^\circ$ mirror fixed to the chamber.
The beams then pass through a quartz window to enter
the drift region of the chamber, which is
filled with a ${\rm CO}_2$/isobutane gas mixture
with one of the following three
proportions: $(85:15)$, $(90:10)$, and $(95:5)$.
The proportions were maintained by mixing
high-purity ${\rm CO}_2$ and isobutane gases
at the corresponding flow rates using a pair
of mass flow controllers.
The total flow rate was kept constant at $100~{\rm cc/min}$
and the temperature and the pressure were continuously
monitored throughout the experiment:
the temperature was $T = 291. \pm 1^\circ{\rm K}$
and $P = 767 \pm 2~{\rm mHg}$. 
The two laser beams ionize the chamber gas molecules
and produce strings of ionization electrons at
two different points in the $z$-direction. 
The spent laser beams leave the chamber through another
quartz window and are reflected 
by another $45^\circ$ mirror to a beam dump.

The produced electron clusters move in a
uniform electric field of $0.6$ to $2.0~{\rm kV/cm}$
provided by a series of ten $1.0~{\rm mm}$-thick copper plates
spaced $5~{\rm mm}$ apart and
connected with a chain of $2.2~{\rm M}\Omega$
resistors (see Fig.~\ref{Fig:chamber}-(b)).
All of these copper plates
have an outer size of $30~{\rm mm} \times 170~{\rm mm}$,
among which the middle eight have
a rectangular hole of $10~{\rm mm} \times 130~{\rm mm}$
that defines the transverse size of the drift region.
The last plate, which is grounded,
has a slit of $3~{\rm mm} \times 130~{\rm mm}$
and separates the drift region and the amplification region behind it.
In the amplification region, 
we have a gold-plated tungsten anode wire 
with a diameter of $30~\mu{\rm m}$ strung $5~{\rm mm}$ away
from and in alignment with the slit.
A typical high voltage applied to the anode wire is
$2.4~{\rm kV}$.
The anode wire is accompanied, in parallel,
by a pair of $120~\mu{\rm m}$-thick
gold-plated Molybdenum wires strung $5~{\rm mm}$ below and above.
On the opposite side of the slit, $5~{\rm mm}$ apart from the anode wire,
there are a series of 27 gold-coated cathode pads 
each being $4.8~{\rm mm}$ in width and $20.0~{\rm mm}$ in height.
The inter-pad gap is $0.2~{\rm mm}$.
In this experiment, however, we used only three of these
cathode pads in the central region to measure
the charge centroids, due to the smallness of Lorentz angles
for our gas mixtures.

\subsection{DAQ System}

\begin{figure}[htb]
\centerline{
\epsfxsize=6cm 
\epsfbox{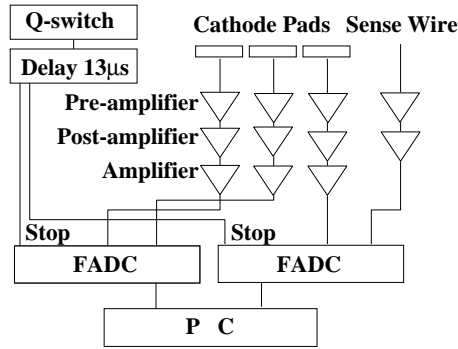}
}
\caption[Fig:BlockDiagram]
{\small \label{Fig:BlockDiagram}
Block diagram for our data acquisition system.
}
\end{figure}

A block diagram of the data acquisition system is shown in
Fig.~\ref{Fig:BlockDiagram}.
%
%
The signals from the three pads and the anode wire
are first amplified by charge-sensitive pre-amplifiers,
and, after passing through a $3~{\rm m}$-long twisted pair
cable, amplified again by post-amplifiers\cite{Ref:amp}.
The cathode pad signals need further amplification
because of their smallness.
We thus put an amplifier with a gain of 8 for each of the three pad signals.
After these amplifications,
the signals are fed into a set of 8-bit 
flash-ADCs\cite{Ref:FADC} 
which have a $500~{\rm MHz}$ sampling frequency
over a $1.6~\mu{\rm s}$ time window.
The flash ADCs convert the analog amplitudes 
from $0~{\rm V}$ to $-1.0~{\rm V}$ 
into a train of digits ranging from 
$255$ to $0$ every $2~{\rm ns}$.
Finally, these flash ADC outputs are recorded
by a PC through a CAMAC system.

%
%

\section{Analysis and Results}
\cleqn
\subsection{Analysis}
\label{Sub:Analysis}

As we discussed in Subsection~\ref{Sub:BasicPrincipleofMes.}, 
Lorentz angle is obtained by measuring 
the transverse distance $\Delta x$ between
the arrival points at the anode wire
of two electron clusters
created in the drift region 
at a distance $\Delta z$ in the electric
field direction.
By means of the cathode pads and
flash ADCs connected to them,
avalanche locations that correspond to
the arrival points of the two electron clusters
($x_1$ and $x_2$)
are encoded as induced-charge distributions.
The $\Delta z$, corresponding to the beam
distance in the electric field direction,
is determined by the optical system and
can be calculated from the time interval ($\Delta t$) 
of the two pulses
at $B = 0$,
together with a pre-measured drift velocity ($v_D^0$)
as $\Delta z = v_D^0 \cdot \Delta t$.

\begin{figure}[htb]
\centerline{
\epsfxsize=6cm 
\epsfbox{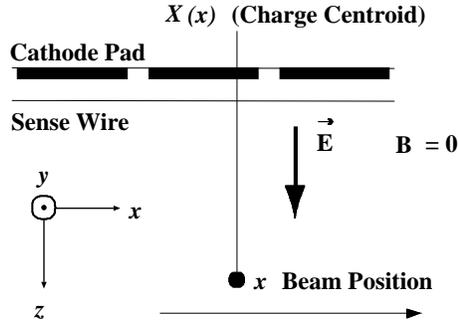}
}
\caption[Fig:calibration]
{\small \label{Fig:calibration}
Calibration scheme. Cathode signals are recorded
with the same readout electronics as used in the
actual Lorentz angle measurement together with the actual
laser beam position ($x$) as obtained from
micrometer readings.
}
\end{figure}

\begin{figure}[htb]
\centerline{
\epsfxsize=7cm 
\epsfbox{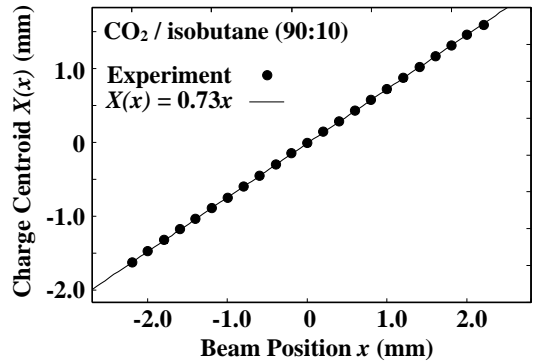}
}
\caption[Fig:calibdat]
{\small \label{Fig:calibdat}
Charge centroid ($X$) as defined in the text as a
function of the injection point ($x$) of
the laser beam, which is used as a calibration curve.
}
\end{figure}

Decoding of the $\Delta x$ information,
on the other hand,
requires a mapping from the charges
on the three cathode pads
to a position in the anode wire direction (the $x$-axis).
It is a common practice to use,
as a measure of this position,
the charge centroid as defined by
\begin{equation}
   X = \frac{\displaystyle \sum_{i=1}^{3} x^i  Q_i }
            {\displaystyle \sum_{i=1}^{3} Q_i}
\label{ChargeCentroid}
\end{equation}
where 
$x^i$ and $Q_i$ are the central position and
the collected charge of the $i$-th cathode pad,
respectively.
The charge centroid ($X$) as defined above,
however, doesn't reproduce the actual
avalanche position ($x$) of an electron cluster,
because of the finite width and number of the cathode pads
used in the calculation.
This necessitates
a calibration curve to map $X$ to $x$.
This map has been obtained by scanning the laser beam
along the $x$-axis with a step of $200~\mu{\rm m}$
by moving the micrometer-adjustable mirror described
in Subsection~\ref{Sub:LaserBeamSystem}
(see Figs.~\ref{Fig:setup} and \ref{Fig:calibration}).
%
%
%
Figure \ref{Fig:calibdat} is an example of
so obtained calibration curves.
%
%
%

\begin{figure}[htb]
\centerline{
\epsfxsize=7cm 
\epsfbox{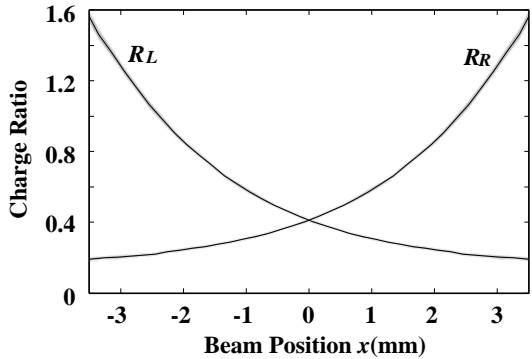}
}
\caption[Fig:calibdat2]
{\small \label{Fig:calibdat2}
Left and right charge ratios as defined in the text as
a function of the injection point ($x$) of the
laser beam, which is used as a calibration curve.
The shaded bands indicate $1$-$\sigma$ error
boundaries.
}
\end{figure}

Another way to map the charge information
to the position is to use the relation between the position and
the ratio of the charge on a side pad to that on the central pad.
There are two ratios available, $R_L \equiv Q_1/Q_2$ 
and $R_R \equiv Q_3/Q_2$.
Their sensitivity to the position varies:
the sensitivity of $R_{L(R)}$
attains its maximum near the boundary of
the left(right) and central pads and
monotonically diminishes towards the other pad boundary.
This is demonstrated in
Fig.~\ref{Fig:calibdat2} which is an example of
the ratio-position relations obtained from
the calibration data we took for the charge centroid method.
%
%
%
Since the two ratios are complementary, we can combine
their information:
\begin{equation}
   x = \frac{\frac{\displaystyle 1}{\displaystyle \sigma_{x(R_L)}^2} 
   	     \cdot x(R_L) +
             \frac{\displaystyle 1}{\displaystyle \sigma_{x(R_R)}^2} 
             \cdot x(R_R) }
            {\frac{\displaystyle 1}{\displaystyle \sigma_{x(R_L)}^2} +
             \frac{\displaystyle 1}{\displaystyle \sigma_{x(R_R)}^2} }.
\end{equation}
Hereafter, we call this method the ratio method.
The ratio method allows us to bypass the calculation
of an intermediate quantity, $X$, and is expected to work
better near the pad boundaries.
The charge centroid method, on the other hand, 
performs better near the center of the central pad.
In this sense, the two methods are complementary.

In either case, the central issue 
is what determines the
calibration curves.
For simplicity, 
let us consider first the case of a single beam.
The induced-charge distribution of the cathode pads
is determined not only by the location,
but also by the shape 
(charge distribution) of the avalanche.
The avalanche shape is affected by
the shape of the electron cluster that initiates
the avalanche, which is in turn depends on
the diffusion that varies with the gas mixture
and the electric field. 
The avalanche shape is also controlled by
the field around the anode wire and
the gas mixture, in particular when
the gas amplification process approaches 
streamer mode operation. 
In order to check the potential dependence on these parameters,
we took calibration data for various electric fields and
for different gas mixtures, 
at several sense-wire high-voltages.
There is no significant 
correlation seen between these parameters
and the calibration curves,
which led us to conclude that the charge centroid
($X$) is a function of a single quantity:
the avalanche location ($x$) in the single beam case.

The determination of $\Delta x$, however,
requires double beams.
We thus have to discuss now the space charge effect of the 
avalanche created by the first electron cluster
on the second one.
There are two possible ways for the first avalanche
to affect the second:
(1) the first avalanche changes the shape of the
second through space charge effects,
and (2) the first avalanche bends the trajectory
of the second electron cluster near the sense wire.
Possibility (1) was eliminated by
comparing the calibration curves of the two pulses:
there was no difference.
In order to test possibility (2),
we changed the intensity of the laser beam
to create the first electron cluster
and monitored $\Delta x$ with no magnetic field,
varying the electric field to change the time interval
of the two avalanche formations.
We saw no significant dependence of $\Delta x$
on the beam intensity nor the electric field $\mathbold E$.
We thus confirmed that
space charge effect is negligible
in our measurement range.

%

\subsection{Results}
\label{Sub:Results}

\begin{figure}[htb]
\centerline{
\epsfxsize=7.0cm 
\epsfbox{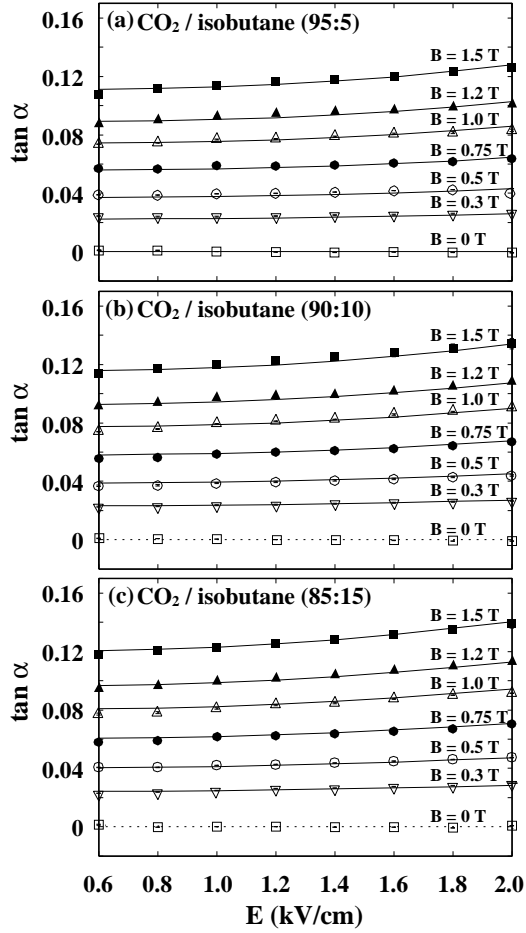}
}
\caption[Fig:results]
{\small \label{Fig:results}
Tangent of the Lorentz angle
($\tan\alpha$) as a function of the electric field
for ${\rm CO}_2$/isobutane mixtures of
(a) (95:5), (b) (90:10), and (c) (85:15).
Smooth curves are GARFIELD/MAGBOLTZ predictions.
}
\end{figure}

Using the $\Delta z$ and the $\Delta x$ values obtained with
the calibration curve discussed above,
we calculated the Lorentz angles 
for different ${\rm CO}_2$/isobutane gas mixtures.
The results are shown in
Figs.~\ref{Fig:results}-(a), -(b), and -(c)
as a function of the electric field
for three mixing ratios:
$(85:15)$, $(90:10)$, and $(95:5)$.
At each electric field value in each figure,
seven points are plotted, corresponding 
to, from bottom to top, seven magnetic field values:
$0.0$, $0.3$, $0.5$, $0.75$, $1.0$, $1.2,$, and $1.5~{\rm T}$,
respectively\footnote[1]{
Knowing that the Lorentz angle and consequently
the transverse displacement $\Delta x$
have to vanish at $B = 0$, 
we averaged 
the zero magnetic field data for each gas mixture
to determine the offset due to the 
transverse misalignment of the two laser beams.
}.
Their numerical values are tabulated in Table.\ref{Table:results}.

%
%
%

The errors contain both statistical and systematic
ones.
The statistical errors have been calculated from the
number of events and found negligible for most cases, 
since a typical resolution for the transverse
distance ($\Delta x$) was as small as
$\sigma_{\Delta x} = 40\mu{\rm m}$.
On the other hand, possible sources of systematic errors
include
\begin{itemize}
\item
	temperature and pressure dependence: negligible,
\item
	laser beam distance: 
	less than $250~\mu {\rm m}$ by a direct survey and 
	was confirmed by the measured time interval of the
	two anode pulses at $B = 0$,
\item
	misalignment of the optical system 
	with respect to the chamber: 
	negligible as long as $\Delta z$ is measured at the\
	points of ionization,
\item
	the electric field value: checked with GARFIELD and
	the field uniformity was confirmed 
	to be much better than $1~\%$,
\item
	non-uniformity of the magnetic field : well below $1~\%$,
\item
	the charge-to-position conversion methods: dominant.
\end{itemize}
\clearpage

%
\begin{table}[htb]
\centerline{
\begin{minipage}[htb]{15cm}
\centerline{
\def\arraystretch{0.9}
{\footnotesize 
\begin{tabular}{|cc|cc|cc|}
\hline
\multicolumn{2}{|c}	{{\footnotesize ${\rm CO}_2$/isobutane (85:15)}} & 
\multicolumn{2}{|c}	{{\footnotesize ${\rm CO}_2$/isobutane (90:10)}} & 
\multicolumn{2}{|c|}	{{\footnotesize ${\rm CO}_2$/isobutane (95:05)}} \cr 
\multicolumn{1}{|c}	{{\footnotesize E(kV/cm)}}	&
\multicolumn{1}{c}	{{\footnotesize tan $\alpha$}}	&
\multicolumn{1}{|c}	{{\footnotesize E(kV/cm)}}	&
\multicolumn{1}{c}	{{\footnotesize tan $\alpha$}}	&
\multicolumn{1}{|c}	{{\footnotesize E(kV/cm)}}	&
\multicolumn{1}{c|}	{{\footnotesize tan $\alpha$}}	\cr
\hline 
\multicolumn{6}{|c|}   {{\footnotesize B = 0.3~T}} \\ 
\hline 
0.6 & $( 2.29 \pm  0.02)\times 10^{-2}$ & 0.6 & $( 2.25 \pm  0.04)\times 10^{-2}$ & 0.6 & $( 2.43 \pm  0.04)\times 10^{-2}$  \\ 
0.8 & $( 2.39 \pm  0.03)\times 10^{-2}$ & 0.8 & $( 2.30 \pm  0.04)\times 10^{-2}$ & 0.8 & $( 2.38 \pm  0.04)\times 10^{-2}$  \\ 
1.0 & $( 2.48 \pm  0.03)\times 10^{-2}$ & 1.0 & $( 2.35 \pm  0.04)\times 10^{-2}$ & 1.0 & $( 2.39 \pm  0.04)\times 10^{-2}$  \\ 
1.2 & $( 2.61 \pm  0.03)\times 10^{-2}$ & 1.2 & $( 2.39 \pm  0.04)\times 10^{-2}$ & 1.2 & $( 2.44 \pm  0.04)\times 10^{-2}$  \\ 
1.4 & $( 2.65 \pm  0.03)\times 10^{-2}$ & 1.4 & $( 2.47 \pm  0.04)\times 10^{-2}$ & 1.4 & $( 2.49 \pm  0.04)\times 10^{-2}$  \\ 
1.6 & $( 2.73 \pm  0.03)\times 10^{-2}$ & 1.6 & $( 2.53 \pm  0.04)\times 10^{-2}$ & 1.6 & $( 2.54 \pm  0.04)\times 10^{-2}$  \\ 
1.8 & $( 2.82 \pm  0.03)\times 10^{-2}$ & 1.8 & $( 2.58 \pm  0.04)\times 10^{-2}$ & 1.8 & $( 2.60 \pm  0.04)\times 10^{-2}$  \\ 
2.0 & $( 2.96 \pm  0.03)\times 10^{-2}$ & 2.0 & $( 2.67 \pm  0.04)\times 10^{-2}$ & 2.0 & $( 2.66 \pm  0.04)\times 10^{-2}$  \\ 
\hline 
\multicolumn{6}{|c|}   {{\footnotesize B = 0.5~T}} \\ 
\hline 
0.6 & $( 4.05 \pm  0.04)\times 10^{-2}$ & 0.6 & $( 3.69 \pm  0.06)\times 10^{-2}$ & 0.6 & $( 3.89 \pm  0.04)\times 10^{-2}$  \\ 
0.8 & $( 4.07 \pm  0.04)\times 10^{-2}$ & 0.8 & $( 3.74 \pm  0.06)\times 10^{-2}$ & 0.8 & $( 3.89 \pm  0.04)\times 10^{-2}$  \\ 
1.0 & $( 4.20 \pm  0.05)\times 10^{-2}$ & 1.0 & $( 3.88 \pm  0.06)\times 10^{-2}$ & 1.0 & $( 3.97 \pm  0.04)\times 10^{-2}$  \\ 
1.2 & $( 4.27 \pm  0.05)\times 10^{-2}$ & 1.2 & $( 3.97 \pm  0.06)\times 10^{-2}$ & 1.2 & $( 3.99 \pm  0.04)\times 10^{-2}$  \\ 
1.4 & $( 4.36 \pm  0.05)\times 10^{-2}$ & 1.4 & $( 4.07 \pm  0.06)\times 10^{-2}$ & 1.4 & $( 4.08 \pm  0.04)\times 10^{-2}$  \\ 
1.6 & $( 4.49 \pm  0.05)\times 10^{-2}$ & 1.6 & $( 4.18 \pm  0.06)\times 10^{-2}$ & 1.6 & $( 4.17 \pm  0.04)\times 10^{-2}$  \\ 
1.8 & $( 4.61 \pm  0.05)\times 10^{-2}$ & 1.8 & $( 4.29 \pm  0.06)\times 10^{-2}$ & 1.8 & $( 4.26 \pm  0.04)\times 10^{-2}$  \\ 
2.0 & $( 4.75 \pm  0.05)\times 10^{-2}$ & 2.0 & $( 4.41 \pm  0.06)\times 10^{-2}$ & 2.0 & $( 4.31 \pm  0.04)\times 10^{-2}$  \\ 
\hline 
\multicolumn{6}{|c|}   {{\footnotesize B = 0.75~T}} \\ 
\hline 
0.6 & $( 5.80 \pm  0.06)\times 10^{-2}$ & 0.6 & $( 5.56 \pm  0.06)\times 10^{-2}$ & 0.6 & $( 5.73 \pm  0.06)\times 10^{-2}$  \\ 
0.8 & $( 5.90 \pm  0.06)\times 10^{-2}$ & 0.8 & $( 5.67 \pm  0.06)\times 10^{-2}$ & 0.8 & $( 5.69 \pm  0.06)\times 10^{-2}$  \\ 
1.0 & $( 6.14 \pm  0.06)\times 10^{-2}$ & 1.0 & $( 5.87 \pm  0.06)\times 10^{-2}$ & 1.0 & $( 5.90 \pm  0.06)\times 10^{-2}$  \\ 
1.2 & $( 6.23 \pm  0.06)\times 10^{-2}$ & 1.2 & $( 6.01 \pm  0.06)\times 10^{-2}$ & 1.2 & $( 5.87 \pm  0.06)\times 10^{-2}$  \\ 
1.4 & $( 6.36 \pm  0.06)\times 10^{-2}$ & 1.4 & $( 6.10 \pm  0.07)\times 10^{-2}$ & 1.4 & $( 5.96 \pm  0.06)\times 10^{-2}$  \\ 
1.6 & $( 6.53 \pm  0.07)\times 10^{-2}$ & 1.6 & $( 6.24 \pm  0.07)\times 10^{-2}$ & 1.6 & $( 6.04 \pm  0.06)\times 10^{-2}$  \\ 
1.8 & $( 6.72 \pm  0.07)\times 10^{-2}$ & 1.8 & $( 6.45 \pm  0.07)\times 10^{-2}$ & 1.8 & $( 6.19 \pm  0.07)\times 10^{-2}$  \\ 
2.0 & $( 7.04 \pm  0.07)\times 10^{-2}$ & 2.0 & $( 6.71 \pm  0.08)\times 10^{-2}$ & 2.0 & $( 6.38 \pm  0.07)\times 10^{-2}$  \\ 
\hline 
\multicolumn{6}{|c|}   {{\footnotesize B = 1.0~T}} \\ 
\hline 
0.6 & $( 7.72 \pm  0.09)\times 10^{-2}$ & 0.6 & $( 7.40 \pm  0.08)\times 10^{-2}$ & 0.6 & $( 7.36 \pm  0.08)\times 10^{-2}$  \\ 
0.8 & $( 7.81 \pm  0.10)\times 10^{-2}$ & 0.8 & $( 7.58 \pm  0.08)\times 10^{-2}$ & 0.8 & $( 7.45 \pm  0.08)\times 10^{-2}$  \\ 
1.0 & $( 8.10 \pm  0.10)\times 10^{-2}$ & 1.0 & $( 7.94 \pm  0.08)\times 10^{-2}$ & 1.0 & $( 7.67 \pm  0.08)\times 10^{-2}$  \\ 
1.2 & $( 8.34 \pm  0.10)\times 10^{-2}$ & 1.2 & $( 8.11 \pm  0.08)\times 10^{-2}$ & 1.2 & $( 7.73 \pm  0.08)\times 10^{-2}$  \\ 
1.4 & $( 8.49 \pm  0.10)\times 10^{-2}$ & 1.4 & $( 8.25 \pm  0.08)\times 10^{-2}$ & 1.4 & $( 7.88 \pm  0.09)\times 10^{-2}$  \\ 
1.6 & $( 8.75 \pm  0.09)\times 10^{-2}$ & 1.6 & $( 8.57 \pm  0.10)\times 10^{-2}$ & 1.6 & $( 8.07 \pm  0.09)\times 10^{-2}$  \\ 
1.8 & $( 9.02 \pm  0.09)\times 10^{-2}$ & 1.8 & $( 8.77 \pm  0.10)\times 10^{-2}$ & 1.8 & $( 8.14 \pm  0.11)\times 10^{-2}$  \\ 
2.0 & $( 9.15 \pm  0.09)\times 10^{-2}$ & 2.0 & $( 8.99 \pm  0.10)\times 10^{-2}$ & 2.0 & $( 8.30 \pm  0.11)\times 10^{-2}$  \\ 
\hline 
\multicolumn{6}{|c|}   {{\footnotesize B = 1.2~T}} \\ 
\hline 
0.6 & $( 9.41 \pm  0.09)\times 10^{-2}$ & 0.6 & $( 9.18 \pm  0.10)\times 10^{-2}$ & 0.6 & $( 8.76 \pm  0.09)\times 10^{-2}$  \\ 
0.8 & $( 9.65 \pm  0.10)\times 10^{-2}$ & 0.8 & $( 9.38 \pm  0.10)\times 10^{-2}$ & 0.8 & $( 9.04 \pm  0.09)\times 10^{-2}$  \\ 
1.0 & $( 9.95 \pm  0.10)\times 10^{-2}$ & 1.0 & $( 9.70 \pm  0.10)\times 10^{-2}$ & 1.0 & $( 9.28 \pm  0.09)\times 10^{-2}$  \\ 
1.2 & $( 1.01 \pm  0.01)\times 10^{-1}$ & 1.2 & $( 9.85 \pm  0.10)\times 10^{-2}$ & 1.2 & $( 9.45 \pm  0.09)\times 10^{-2}$  \\ 
1.4 & $( 1.04 \pm  0.01)\times 10^{-1}$ & 1.4 & $( 9.91 \pm  0.11)\times 10^{-2}$ & 1.4 & $( 9.61 \pm  0.10)\times 10^{-2}$  \\ 
1.6 & $( 1.07 \pm  0.01)\times 10^{-1}$ & 1.6 & $( 1.02 \pm  0.01)\times 10^{-1}$ & 1.6 & $( 9.71 \pm  0.10)\times 10^{-2}$  \\ 
1.8 & $( 1.10 \pm  0.01)\times 10^{-1}$ & 1.8 & $( 1.05 \pm  0.01)\times 10^{-1}$ & 1.8 & $( 9.91 \pm  0.10)\times 10^{-2}$  \\ 
2.0 & $( 1.13 \pm  0.01)\times 10^{-1}$ & 2.0 & $( 1.08 \pm  0.01)\times 10^{-1}$ & 2.0 & $( 1.01 \pm  0.01)\times 10^{-1}$  \\ 
\hline 
\multicolumn{6}{|c|}   {{\footnotesize B = 1.5~T}} \\ 
\hline 
0.6 & $( 1.18 \pm  0.02)\times 10^{-1}$ & 0.6 & $( 1.14 \pm  0.02)\times 10^{-1}$ & 0.6 & $( 1.08 \pm  0.01)\times 10^{-1}$  \\ 
0.8 & $( 1.20 \pm  0.01)\times 10^{-1}$ & 0.8 & $( 1.17 \pm  0.02)\times 10^{-1}$ & 0.8 & $( 1.12 \pm  0.01)\times 10^{-1}$  \\ 
1.0 & $( 1.22 \pm  0.02)\times 10^{-1}$ & 1.0 & $( 1.20 \pm  0.02)\times 10^{-1}$ & 1.0 & $( 1.14 \pm  0.01)\times 10^{-1}$  \\ 
1.2 & $( 1.25 \pm  0.02)\times 10^{-1}$ & 1.2 & $( 1.23 \pm  0.02)\times 10^{-1}$ & 1.2 & $( 1.17 \pm  0.01)\times 10^{-1}$  \\ 
1.4 & $( 1.28 \pm  0.02)\times 10^{-1}$ & 1.4 & $( 1.25 \pm  0.02)\times 10^{-1}$ & 1.4 & $( 1.19 \pm  0.01)\times 10^{-1}$  \\ 
1.6 & $( 1.32 \pm  0.02)\times 10^{-1}$ & 1.6 & $( 1.28 \pm  0.03)\times 10^{-1}$ & 1.6 & $( 1.21 \pm  0.01)\times 10^{-1}$  \\ 
1.8 & $( 1.35 \pm  0.02)\times 10^{-1}$ & 1.8 & $( 1.31 \pm  0.03)\times 10^{-1}$ & 1.8 & $( 1.23 \pm  0.01)\times 10^{-1}$  \\ 
2.0 & $( 1.39 \pm  0.03)\times 10^{-1}$ & 2.0 & $( 1.34 \pm  0.04)\times 10^{-1}$ & 2.0 & $( 1.26 \pm  0.02)\times 10^{-1}$  \\ 
\hline 
\end{tabular}
}
}
\caption[Table:results]
{\small \label{Table:results}
$\tan\alpha$ data
}
\end{minipage}
}
\end{table}

\clearpage

As we discussed at the beginning of this section,
there are two methods to convert cathode pad charge
information to position information: the charge centroid method
and the ratio method.
The sensitivity of the two methods differ depending on
the location of the center of the induced charge.
In general the ratio method works better when the
charge center is near the pad boundaries, 
while the charge centroid method is expected to perform
better near the center of the central pad.
Because of this, we tried both methods to calculate $\Delta x$.
In some cases, we found some discrepancies
between them.
Since there was no a priori reason to prefer
particular one to the other,
we decided to take the average of the two and put
the half of the discrepancy as a systematic error on each point.

Curves in Figs.~\ref{Fig:results}-(a), -(b), and -(c)
are predictions obtained with MAGBOLTZ-1 (version 1.16) through
its GARFIELD interface,
although its accuracy for the velocity distribution of electrons
is known to be limited under certain circumstances\cite{Ref:robson}.
The loss of accuracy is caused by a decomposition of the
velocity distribution function in Legendre polynomials,
in which the lowest two or three terms are retained in the calculation.
Therefore results given by the program may not be precise
enough when the velocity distribution deviates far from
isotropy or it has no axial symmetry as in the case
of crossed electric and magnetic fields.
On the other hand, our experimental condition seems to be
favorable for application of MAGBOLTZ-1:
electrons in ${\rm CO}_2$-based gas mixtures under a low electric
field are nearly thermal and the axial symmetry of the velocity
distribution holds to a good extent,
because their Lorentz angles are small even in a high
magnetic field.
To confirm this we ran the Monte Carlo version 
(MAGBOLTZ-2, version 2.2\cite{Ref:MAGBOLTZ2}),
which is free from the problems stated above
though time-consuming,
to simulate Lorentz angles for several electric and magnetic
field combinations.
The results were found to be consistent with those
obtained with MAGBOLTZ-1.
 
Our results are in good agreement
with the MAGBOLTZ-1 predictions.

%
%

\section{Discussion}
\label{Sec:discussion}
\cleqn

\subsection{Magnetic Deflection Coefficient \protect\boldmath$\psi$}
\label{Sub:psi}

\begin{figure}[htb]
\centerline{
\epsfxsize=7cm 
\epsfbox{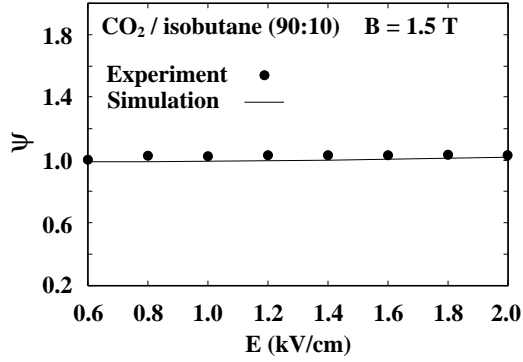}
}
\caption[Fig:psi]{\small \label{Fig:psi}
Magnetic deflection coefficient at $B = 1.5$T
as a function of the electric field
for the ${\rm CO}_2$/isobutane(90:10) mixture.
}
\end{figure}

\begin{figure}[htb]
\centerline{
\epsfxsize=7cm 
\epsfbox{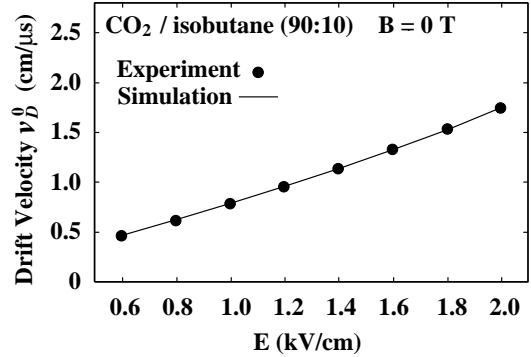}
}
\caption[Fig:velocity]{\small \label{Fig:velocity}
Drift velocity at $B = 0$
as a function of the electric field
for the ${\rm CO}_2$/isobutane(90:10) mixture.
}
\end{figure}

\begin{figure}[htb]
\centerline{
\epsfxsize=7cm 
\epsfbox{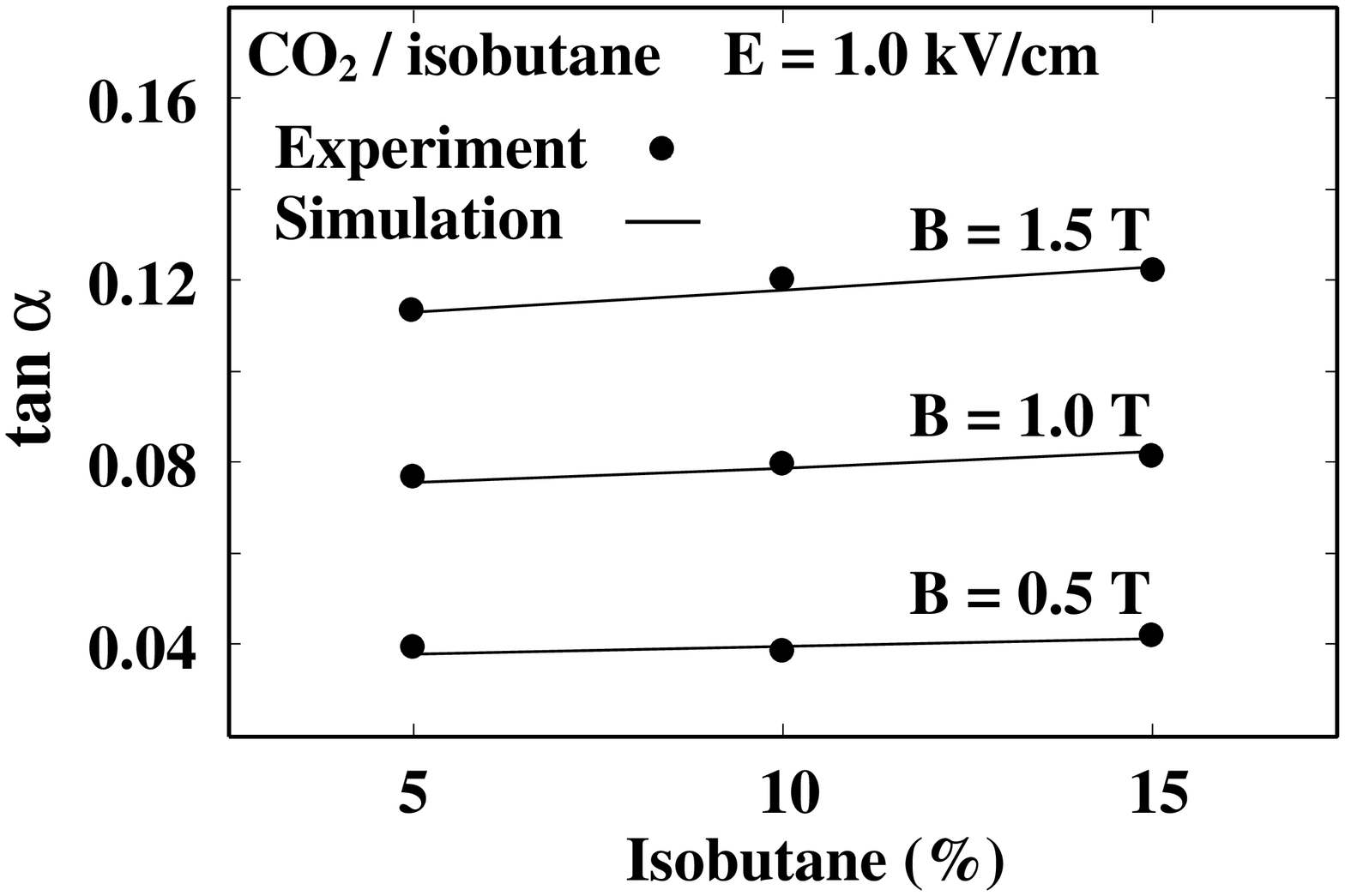}
}
\caption[Fig:gasmixtures]
{\small \label{Fig:gasmixtures}
Isobutane concentration dependence of $\tan\alpha$
at $E = 1.0~$kV/cm.
The solid lines are GARFIELD/MAGBOLTZ predictions.
}
\end{figure}

	We calculated the magnetic deflection coefficient ($\psi$)
from the measured Lorentz angles and the drift velocities obtained 
without magnetic field.
Fig.~\ref{Fig:psi} shows the resultant $\psi$ as a function of 
electric field strength
for the CO$_2$/isobutane(90:10) mixture at $B = 1.5$~T,
while Fig.~\ref{Fig:velocity} shows the drift velocity in the 
absence of magnetic field($v_D^0$).
%
The drift velocity is also listed in Table~\ref{Table:velocity}.
\begin{table}[htb]
\centerline{
{\footnotesize 
\begin{tabular}{|cc|}
\hline
\multicolumn{1}{|c}	{{\footnotesize E~(kV/cm)}}	&
\multicolumn{1}{c|}	{{\footnotesize $v_D^0$~(cm/$\mu$s)}}	\cr
\hline
\def\arraystretch{0.9}
0.6 & $0.453 \pm 0.002$ \\ 
0.8 & $0.607 \pm 0.001$ \\ 
1.0 & $0.780 \pm 0.001$ \\ 
1.2 & $0.950 \pm 0.001$ \\ 
1.4 & $1.132 \pm 0.001$ \\ 
1.6 & $1.323 \pm 0.001$ \\ 
1.8 & $1.524 \pm 0.004$ \\ 
2.0 & $1.735 \pm 0.004$ \\ 
\hline
\end{tabular}
}
}
\caption[Table:velocity]
{\small \label{Table:velocity}
Drift velocity data for ${\rm CO}_2$/isobutane (90:10) at $B = 0$~T.
}
\end{table}
The values of $\psi$ were found to be close to unity within $\pm$ 5\%
for the whole range of the applied electric and magnetic fields
and for all the gas mixtures used.
The gas dependence of the Lorentz angle is shown in Fig.~\ref{Fig:gasmixtures}.
The observed increase of the Lorentz angle with isobutane concentration
is consistent with the increase of drift velocity and with $\psi = 1$.
Why are the values of $\psi$ so close to unity?
%

	As discussed in Subsection~\ref{Sub:TheoreticalReview}, 
$\psi = 1$ is a straightforward consequence
of a simplified model of the force acting on a drifting electron.
The model is based on naive expectation  for the average
force from gas molecules to be anti-parallel with the drift velocity
(Eq.~\ref{Eq:avat})
and a crude assumption that the average time interval between collisions ($\tau$)
is not affected by the presence of a magnetic field.

	In general, however, the average stochastic force, namely the average momentum 
transfer per unit time from the gas molecules to an electron,
is given by
\begin{eqnarray}
\label{Eq:Fcoll}
 \left< \mathbold{F}_{\rm coll} \right> 
    &\equiv& m \left< \mathbold{A}(t) \right>  ~=~ 
    - \int \nu_m (v)\, m \mathbold{v} f(\mathbold{v})\,{\rm d}\mathbold{v} \cr
    &\simeq& - m\nu_m(v_0) \int \mathbold{v} f(\mathbold{v})\,{\rm d}\mathbold{v}
    - m \left. \frac{{\rm d}\nu_m}{{\rm d}v} \right|_{v=v_0} \int ({v-v_0})
    \mathbold{v} f(\mathbold{v})\,{\rm d}\mathbold{v},
\end{eqnarray}
where $\nu_m$ is momentum-transfer collision frequency\footnote[1]{
The momentum transfer collision frequency is
given by
\begin{eqnarray*}
  \nu_m(|\mathbold{v}|) &\equiv& N |\mathbold{v}| \int (1-\cos\theta) {\rm d}\sigma \cr
        & = & N |\mathbold{v}| \sigma \, (1-\overline{\cos\theta}),
\end{eqnarray*}
where $N$ is the density of gas molecules, ${\rm d}\sigma$
is the differential cross section, and $\theta$
is the scattering angle measured from $\mathbold{v}$.
},
$f(\mathbold{v})$ represents the normalized random velocity 
distribution of electrons,
and $v_0$ is an appropriate average value of $|\mathbold{v}|$.
Thus $ \left< \mathbold{F}_{\rm coll} \right>$
is not necessarily anti-parallel with the drift velocity
($\mathbold{v}_{\it D} 
= {\displaystyle\int \mathbold{v}\it{f}(\mathbold{v})\,{\rm d}\mathbold{v}}$)
unless $f(\mathbold{v})$ has axial symmetry with respect to the drift direction
(as in the case of no magnetic field)
or $\nu_m$ is independent of $|\mathbold{v}|$
\footnote[2]{In fact the axial symmetry of $f(\mathbold{v})$ is retained
even under a magnetic field when $\nu_m$ is constant\cite{Ref:Huxley}.}. 
Furthermore $\tau$ ( $\simeq 1/\nu_m(v_0)$) does depend on $B$ since the magnetic field 
modifies the energy distribution of electrons.
Therefore $\psi$ is not in general expected to be unity.

	It may be possible to make the simplified model a little bit more realistic
by adding the component of $ \left< \mathbold{F}_{\rm coll} \right>$
perpendicular to $\mathbold{v}_{\it D}$
($ \left< \mathbold{F}_{\rm coll} \right>_\perp$)
to the fundamental equation (\ref{Eq:taeqm})
and by assuming $\tau$ to be a function of $E/N$ and $B/N$,
where $N$ is the density of gas molecules.
$ \left< \mathbold{F}_{\rm coll} \right>_\perp$
may be expressed as
$-e\,K\,\mathbold{v}_D \times \mathbold{B}$ with $K$,
again a function of both 
$E/N$ and $B/N$.
The relation (\ref{Eq:tana-psi=1}) is then corrected to be
\begin{eqnarray}
   \tan \alpha 
 &=& ( 1 + K )\frac{eB}{m} \, \tau  \cr 
 &=& ( 1 + K ) \left (\frac{\tau}{\tau_0} \right) \left( \frac{B}{E} \right) v_D^0 \,, 
\end{eqnarray}
where $\tau = \tau \, (E/N,B/N)$ and $\tau_0 = \tau \,(E/N,0)$.
Therefore the magnetic deflection coefficient is defined by 
\begin{equation}
 \psi = ( 1 + K ) \, \frac{\tau}{\tau_0}\,.
\end{equation}
The sign of $K$ is positive in most cases.
Therefore $ \left< \mathbold{F}_{\rm coll} \right>_\perp$
shows up as an increase of the apparent strength of
$\mathbold{B}$ in Eq.(\ref{Eq:taeqm}) 
and consequently contributes to make $\psi$ greater than unity.
It is worth mentioning that even in the case where $B$ is small enough
to assure that $\tau / \tau_0 = 1$, $\psi$ can be greater than unity 
because of the existence of positive $K$ 
\footnote[3]{
The $B/N$-dependence of the function $K$ and $\tau$,
and therefore $\psi$ is expected to be virtually diminished 
if the effective reduction of $E/N$ is 
taken into account\cite{Ref:Becker2}.  }.

	In the gas mixtures tested and under the applied electric field
($\le$ 2 kV/cm), electrons are nearly thermal, i.e.,
the velocity distribution is close to Maxwellian,
and $\nu_m$ is
fairly constant over the main portion of the electron velocity (energy)
distribution\cite{Ref:Hake-Phelps}.
Under these conditions, the use of the second line of Eq.(\ref{Eq:Fcoll}) 
leads to an approximate expression for the function $K$:
\begin{eqnarray}
 K ~\sim~ \cos^2\alpha \,\left( \frac{\nu^{\scriptstyle\prime}_m(v_0)}{\nu_m(v_0)} \right)^2
\, \left< (\Delta v)^2 \right>
\end{eqnarray}
with $ \left< (\Delta v)^2 \right> \equiv {\displaystyle\int (v-v_0)^2 f_0(v)\,{\rm d}v}$, 
where $\nu^{\scriptstyle\prime}_m(v_0) \equiv 
 \left. \frac{{\rm d}\nu_m}{{\rm d}v} \right|_{v=v_0}$
and $f_0(v) \equiv {\displaystyle\int v^2 f(\mathbold{v})\,{\rm d}\Omega}$.
$K$ is certainly positive and estimated to be much smaller than one because of
small $\nu^{\scriptstyle\prime}_m(v_0)$, large $\nu_m(v_0)$ and
the narrow energy distribution.
Besides, the variation of $\tau$ and
$\left< (\Delta v)^2 \right>$
caused by the magnetic field is
insignificant since  the energy distribution of electrons is modified little.
These facts explain why $\psi$ is always close to unity
in our measurements.

	It should be noted that the above argument is justified
for relatively low  $E/N$ and $\psi$ may deviate from unity under higher electric fields 
as in the case of argon-based gas mixtures
\cite{Ref:Becker2}\cite{Ref:Breskin}.

\subsection{Extrapolation to 2\,T}

\begin{figure}[htb]
\centerline{
\epsfxsize=7cm 
\epsfbox{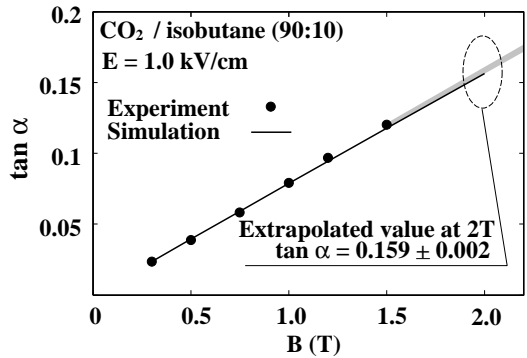}
}
\caption[Fig:extra90E10garf]
{\small \label{Fig:extra90E10garf}
$\tan\alpha$ at $E = 1.0$~kV/cm plotted 
against the magnetic field for the 
${\rm CO}_2$/isobutane(90:10) mixture.
The shaded band above $B = 1.5$~T is the $1$-$\sigma$
bound for the straight-line extrapolation,
while the solid line is the GARFIELD/MAGBOLTZ simulation.
}
\end{figure}

The current design of the JLC-CDC 
assumes operation 
under a magnetic field of $2.0~{\rm T}$.
Figure~\ref{Fig:extra90E10garf} plots
the Lorentz angle as a function of the magnetic field.
As is expected from Eq.(\ref{Eq:tanA}),
the Lorentz angle is proportional to the
magnetic field as long as $\psi = 1$.
We thus fit a straight line passing through the origin
to the data points below $1.5~{\rm T}$
and extrapolate the line to $2.0~{\rm T}$,
in order to estimate the Lorentz angle for
the JLC-CDC.
At $E = 1~{\rm kV/cm}$ and $B = 2~{\rm T}$ 
the extrapolated Lorentz angle is
$\tan\alpha = 0.159 \pm 0.002$ for the ${\rm CO}_2$/isobutane(90:10)
mixture.
The shaded band above $1.5~{\rm T}$
indicates $1$-$\sigma$ extrapolation error interval.
The dotted line in the figure is the prediction
of GARFIELD/MAGBOLTZ which is consistent with 
the extrapolation.
The JLC cell design shown in 
Fig.~\ref{Fig:drift-magn} is thus justified.


\subsection{Possibility of Higher Magnetic Field}

Motivated mainly by recent studies
of beam-induced background\cite{Ref:beamBG},
possibility of higher magnetic field
is now under serious considerations.
Our current cell design allows a magnetic
field up to about $3~{\rm T}$.
Since GARFIELD/MAGBOLTZ reproduces
our Lorentz angle data very well,
it is plausible that it continues to work
well at around $3~{\rm T}$, too.
It is, however, desirable to confirm
this experimentally.
We are thus planning to measure Lorentz angles
at higher magnetic fields.

\section{Conclusions}
\cleqn

We have developed a new Lorentz angle measurement system
for cool gas mixtures.
The measurement system features a laser beam system
providing two simultaneous beams to produce primary electrons
and a drift chamber with flash ADCs to read their signals.
The use of the two simultaneous laser beams
and the flash ADCs
significantly reduced
systematic errors and allowed
measurements of small Lorentz angles
expected for cool gas mixtures such as
${\rm CO}_2$-based mixtures.	

Using this new system, we have
measured Lorentz angles for
${\rm CO}_2$/isobutane gas mixtures
with different proportions:
(95:5, 90:10, and 85:15),
varying drift field from
$0.6$ to $2.0~{\rm kV/cm}$
and magnetic field up to $1.5~{\rm T}$.
The results of the measurement are
in good agreement with GARFIELD/MAGBOLTZ simulations.
Our data confirmed
the validity of the assumption of
the magnetic deflection coefficient being unity
in our measurement range.
The mixing ratio dependence of the Lorentz angle can thus
be understood simply by that of the drift velocity.

We used our data to estimate the
Lorentz angle for the current JLC-CDC cell design.
At $E = 1~{\rm kV/cm}$ and $B = 2~{\rm T}$ 
the extrapolated Lorentz angle is
$\tan\alpha = 0.159 \pm 0.002$ for the ${\rm CO}_2$/isobutane(90:10)
mixture,
confirming the validity of the current design.

\section*{Acknowledgments}

The authors would like to thank all the members of 
the JLC physics working group.
In particular, they are very grateful to
Y.~Asano, T.~Matsui,
and S.~Iwata for continuous encouragements and supports.
This work was supported in part by the JSPS Japanese-German
Cooperation Program.


\end{document}